\documentclass[prl,twocolumn,floatfix,showpacs,superscriptaddress]{revtex4}
\usepackage{amsmath}
\usepackage[dvips]{graphicx}% Include figure files

\newcommand{\eq}[2]
{\begin{equation}
    #1
    \label{#2}
  \end{equation}}

\newcommand{\eqalign}[2]
{
  \begin{align*}
    #1
    \label{#2}
    \end{align}
}

\def\bcen{\begin{center}}
\def\ecen{\end{center}}

        \def\o{\omega}

\def\eg{\mbox{\it e.g. }}
\def\ie{\mbox{\it i.e. }}
\def\=={\equiv}

\def\qed{\raise1pt\hbox{\vrule height5pt width5pt depth0pt}}

\def\cG0{{\cal G}_0}
\def\cG{{\cal G}}

\def\o{\omega}

\def\=={\equiv}

\begin{document}

\title{Mottness scenario for non-Fermi liquid behavior in the Periodic Anderson Model within Dynamical Mean Field Theory}
\author{A. Amaricci}
\affiliation{Laboratoire de Physique des Solides, CNRS-UMR8502, Universit\'e de Paris-Sud,
Orsay 91405, France.}
\affiliation{Dipartimento di Fisica, Universit\`a di Roma ``Tor Vergata'',
Roma 00133, Italy.}
\author{G. Sordi}
\affiliation{Laboratoire de Physique des Solides, CNRS-UMR8502, Universit\'e de Paris-Sud,
Orsay 91405, France.}
\author{M.J. Rozenberg}
\affiliation{Laboratoire de Physique des Solides, CNRS-UMR8502, Universit\'e de Paris-Sud,
Orsay 91405, France.}
\affiliation{Departamento de F\'{\i}sica, FCEN, Universidad de Buenos Aires,
Ciudad Universitaria Pab.I, Buenos Aires (1428), Argentina.}

\date{today}
\begin{abstract}
We study the 
Mott metal-insulator transition in the Periodic Anderson Model within Dynamical
Mean Field Theory (DMFT). Near the quantum transition, we find a
non-Fermi liquid metallic state down to a vanishing temperature
scale.
We identify the origin of the non-Fermi liquid behavior as due to
magnetic scattering of the doped carriers by the localized moments. 
The non-Fermi liquid state can be tuned by either doping or external
magnetic field.
Our results show that the coupling to spatial magnetic 
fluctuations (absent in DMFT) is not a prerequisite to realize a
non-Fermi liquid scenario for heavy fermion systems.
\end{abstract}


\maketitle
The theoretical understanding of the breakdown of the Fermi liquid paradigm
observed in high $T_c$ superconductors and heavy fermions systems remains
one of the open challenges in strongly correlated physics.
These systems show metallic phases with anomalous properties that cannot be
accounted for by the Fermi liquid theory, which provides an adequate description
of the electronic state of ordinary metals. The reason is intimately
related to the strong correlation effects, originated
in the localized nature of the $d$ and $f$ orbitals of the
experimental compounds \cite{stewart}.
Different ideas have been proposed over the years to try
to explain the origin of the non-Fermi liquid (NFL) states, without an absolute
consensus so far. However, many of these ideas share a common feature, namely that
the central ingredient is the proximity to a quantum phase transition (QPT),
or a quantum critical point (QCP) \cite{sachdev}.
In that scenario the breakdown of the Fermi liquid occurs in the neighborhood
of a $T=0$ transition between an ordered phase (\eg antiferromagnetic)
and a paramagnetic one. There, the fluctuations of the order
parameter that couple to the electrons are strongest, and are viewed as
the origin for the NFL state. Among those approaches we can mention
the Hertz-Millis theory, where the paramagnons of the ordered
phase ``dress'' the conduction electrons to produce the NFL features
\cite{millis}.
Another approach is the local quantum critical theory \cite{qimiao,naturephys},
which does not consider the electrons as ``bystanders'' but
emphasizes their role in the screening of the local magnetic moments, via
the celebrated Kondo effect.
There, the competition between the tendency to formation of local singlets
and the long wavelength magnetic fluctuations are to be considered on
equal footing. This has been achieved by the formulation of an extension to
the Dynamical Mean Field Theory (DMFT) approach \cite{rmp},
called EDMFT \cite{edmft1,edmft2}.
DMFT has proven to be a very useful technique to study strongly correlated
electron systems when the main physical effects are local \cite{rmp}.
However, it is also recognized that this method lacks a proper
description of the spatial magnetic fluctuations which are
usually considered a crucial ingredient for the realization of a NFL state.
This shortcoming is cured in EDMFT with the incorporation of a bosonic
component to the effective mean field.
Though that approach has provided useful insight into the physics of the
problem \cite{edmft1,edmft2,sk,edmft},
the solution of its mean field equations is more
difficult and often requires some simplifying assumptions.
We should also mention that a different and original
approach to quantum criticality that
goes beyond the Ginzburg-Landau theory has also been proposed
\cite{senthil}.

In the present work we shall show that, contrary to conventional
expectations, a non-Fermi liquid state is readily obtained from the
DMFT solution of the canonical model for the study of heavy fermion
systems, namely, the periodic Anderson model (PAM) \cite{jarrell}.
We shall show that, similarly to other approaches this novel NFL state
is located in the neighborhood of a QPT, but unlike the standard scenario
described before, the relevant quantum transition here is a Mott transition. 
%between two magnetically disordered phases, one metallic and one a correlation
%driven Mott insulator.
Thus, the present study sheds a different light onto the problem,
showing that the coupling to long wavelength magnetic fluctuations
(which are absent in DMFT) is not a prerequisite for the realization
of a NFL scenario that captures some features observed in heavy fermion systems. 
{\it Local} temporal magnetic fluctuations alone
can provide sufficient scattering to produce an incoherent
metallic state.
Two recent works have also reported non-Fermi liquid states
in models that generalize the PAM by introducing a finite bandwidth
to the localized orbitals \cite{pepin,bmg}.
However, those approaches fundamentally
differ from the present one, since they view the NFL state as the result
of an orbital selective Mott transition, which is absent in the PAM.
Thus, our study shows that the PAM, solved within DMFT (\ie
in the limit of infinite dimensions), may be considered
as a ``bare bones'' or minimal approach
that can qualitatively capture a physical scenario found in some NFL heavy fermion
systems.
%Among our main results we find that in the NFL state the local spin
%susceptibility of the heavy electrons is strongly enhanced, while the
%lifetime of the conduction electrons remains finite down to the
%low $T$ limit.
%Nevertheless, and rather surprisingly, the density of
%states (DOS) of the electrons near the Fermi energy displays a
%quasiparticle-like peak with the peak-dip-hump shape,
%reminiscent to what is observed in photoemission studies of cuprates \cite{dessau}.
%In addition, and similar to experiments in heavy fermion
%compounds \cite{stewart}, the NFL state crosses over towards a normal Fermi
%liquid upon heavier doping or the application of an external
%magnetic field.

The periodic Anderson model Hamiltonian reads,
\eq{
\begin{split}
H=&-t\sum_{<ij>\sigma} (p^+_{i\sigma}p_{j\sigma} + p^+_{j\sigma}p_{i\sigma}) +
\left(\epsilon_p - \mu \right) \sum_{i\sigma} p^+_{i\sigma}p_{i\sigma} \\
 &+(\epsilon_d-\mu)\sum_{i\sigma} d^+_{i\sigma}d_{i\sigma}
 +t_{pd} \sum_{i\sigma}\left( d^+_{i\sigma}p_{i\sigma} +p^+_{i\sigma}d_{i\sigma} \right) \\
&+U\sum_i  \left(n_{di\uparrow}-\tfrac{1}{2}\right)\left(n_{di\downarrow}-\tfrac{1}{2}\right)
\end{split}}{PAMHam}
\noindent
where $p_{i\sigma}$ and $p^+_{i\sigma}$ are the destruction and creation operators for
the electrons at $p$ orbitals with energy $\epsilon_p$ and
hopping parameter $t$. $d_{i\sigma}$ and $d^+_{i\sigma}$ are the respective operators
of the non-dispersive $d$-electron orbitals with energy $\epsilon_d$, that
we fix equal to 0 without loss of generality \cite{note}.
The $p$ and $d$ orbitals are hybridized with an amplitude $t_{pd}$,
and the electron correlations are introduced by the Coulomb interaction
$U$ on the $d$ sites only.
For simplicity we consider the model defined on a Bethe lattice
that gives a semicircular DOS for the $p$-electron band
(at $t_{pd}$ and $U=0$)\cite{rmp}. Its half-bandwidth is $D=2t=1$ and sets
the units of the problem.
The charge transfer energy is defined as $\Delta = \epsilon_d - \epsilon_p$,
and $\mu$ is the chemical potential.
At $U=0$ and large $\Delta$, the hybridization parameter $t_{pd}$ permits
the delocalization of the $d$ electrons,
as a narrow band forms at the Fermi energy with
bandwidth $\sim t_{pd}^2/\Delta$.
However, when both $U$ and $\Delta$ are large, and the narrow band is half-filled, the
periodic Anderson model can describe two different correlated
insulator states: a charge transfer insulator for $\Delta < U$,
and a Mott-Hubbard one for $U < \Delta$.

The doping driven metal-insulator transition (MIT) in the
paramagnetic Mott-Hubbard insulator was the focus
of our recent study \cite{sordi}. The main finding was a
qualitatively different scenario for the electron or hole driven transitions.
In the former case the MIT was expectedly similar to
the first order transition of the well studied one band Hubbard model \cite{rmp}.
However, in the latter case,  an intriguing second order transition was found.
Here, we shall study in detail the unexpected nature of this Mott transition.
%and show that the metal that is obtained by hole doping
%is a novel non-Fermi liquid state.
%Similarly as in experiments, this state eventually crosses over to a
%normal Fermi liquid when the doping is further increased or when
%an external magnetic field is applied.

Within DMFT the PAM can be mapped onto a quantum impurity
problem subject to a
self-consistency condition \cite{gks}.
Using the cavity method \cite{rmp} one obtains the effective action for the
quantum impurity model:
\eq{
\begin{split}
{\rm S_{\rm eff}}=&
-\int_0^{\beta}d\tau \int_0^\beta d\tau' \sum_{\sigma} d^+_{\sigma}(\tau)
{\cal G}^{-1}_0(\tau - \tau')d_{\sigma}(\tau') \\
&+U\int_0^{\beta} d\tau \left[n_{d\uparrow}(\tau)-\tfrac{1}{2}\right]\left[n_{d\downarrow}(\tau)
-\tfrac{1}{2}\right]
%+{\rm const.}
\end{split}
}{Seff}
where $d_{\sigma}$ and $d^+_{\sigma}$ destroy and create a $d$ electron on an
arbitrarily chosen site and $n_{d\sigma}$ is the occupation operator.
${\cal G}_0$ is the function describing the properties
of the dynamical bath of the site, which is subject to the self-consistent
condition,
\eq{
{\cal G}_0^{-1}(i\omega_n)=
i\omega_n+\mu-\epsilon_d-\frac{{{t^2_{pd}}}}{i\omega_n+\mu-\epsilon_p-t^2G_{pp}(i\omega_n)} }
{self}
The $p$-electron Green function $G_{pp}$,
%containing the information of the ``effective medium'',
is  written in terms of the local self-energy $\Sigma_{pp}$ and the non interacting density
of states $\rho^0$ as:
\eq{
G_{pp}(i\omega_n)=\int{ d\epsilon \frac{\rho^0(\epsilon)}
{i\omega_n +\mu -\epsilon_p -\Sigma_{pp}(i\o_n) - \epsilon}}\
}{gpp}
where $\epsilon$ is the single particle energy, and $\Sigma_{pp}$ is obtained
from the quantum impurity model self-energy $\Sigma(i\omega_n)$
of Eq.~\ref{Seff} \cite{rmp,gks},
\eq{
\Sigma_{pp}(i\o_n)=\frac{t^2_{pd}}
{i\o_n+\mu-\epsilon_d-\Sigma(i\o_n)}
}{sigmapp}
%We will see in the following that this is the
%proper quantity to look at in the small hole
%doped regime of PAM, being the d-electrons essentially ``freezed''.

To solve these equations we use two, in principle
exact, numerical methods: Quantum Monte Carlo (QMC) \cite{hf} and Exact
Diagonalization with the Density
Matrix Renormalization Group (ED-DMRG)\cite{dmrg}.
%DMRG algorithm was introduced as impurity solver for DMFT in [cite GHR04]. It permits
%an improvements of Exact Diagonalization solution, making no a priori approximation.
%We will not review in detalis these methods here for reason of space saving, reminding
%the reader to the very well documented literature [ref.:QMC,ED,DMRG].
The former is a finite temperature calculation and is exact in a statistical sense,
while the latter is a $T=0$ method that relies on diagonalization of large finite clusters.
The comparison of the results obtained from these methods allows for a non-trivial benchmark
of the numerical results.
In our simulations we typically perform over $10^6$ Monte Carlo
sweeps in order to minimize the statistical errors.

\begin{figure}%[!ht]
\centering
\includegraphics[width=9cm,clip]{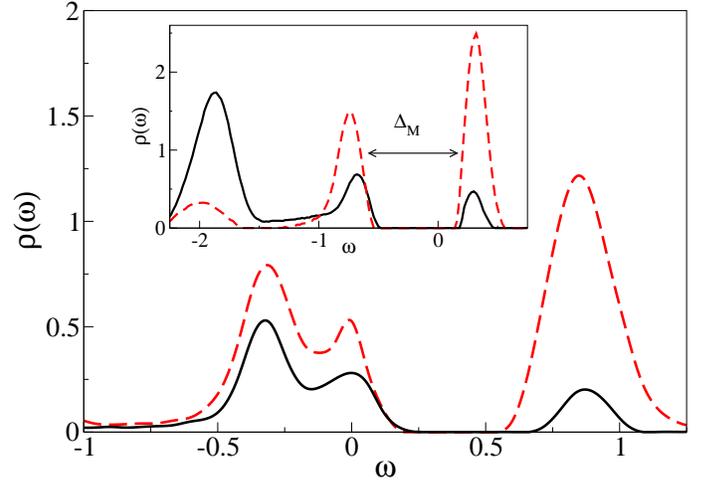}
\caption{
Density of states for $p$ and $d$ electrons
%0.015625
(solid and dashed line respectively) from QMC data at T=1/128. The doping is $\delta=0.049$,
$\mu=0.53$, $\Delta=1$, $U=2$ and $t_{pd}=0.9$. The analytic continuation is
done using maximum entropy method.
%{Left inset}: comparison of DOS from ED-DMRG and QMC data (thin and thick line respectively).
{Inset}: DOS of $p$ and $d$ electrons in the Mott-Hubbard insulating phase.
The chemical potential is $\mu=1.08$ and the occupation is $n=n_d+n_p=3$, with $n_d=1+\nu$, 
$n_p=2-\nu$ and $\nu=0.14$.  $\Delta_{\rm M}$ denotes the Mott insulating gap.
}
\label{fig1}
\end{figure}

\begin{figure}%[!ht]
\centering
\includegraphics[width=9cm,clip]{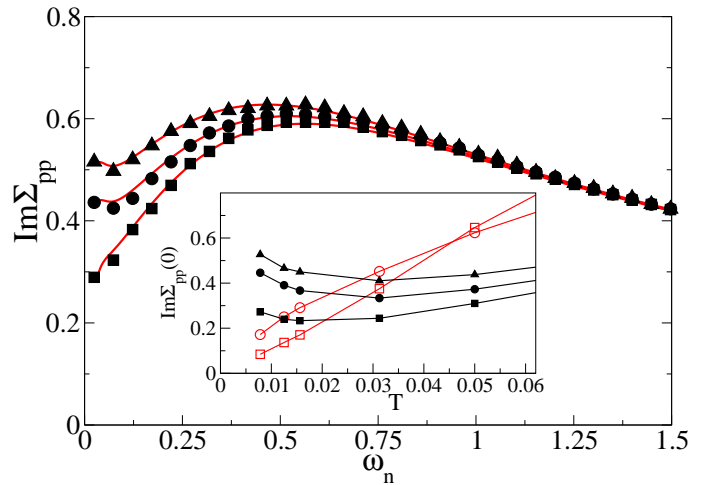}
\caption
{
Imaginary parts of the $p-$electron self-energy Im$\Sigma_{pp}(i\omega_n)$
in the NFL state, from QMC at T=1/128. The hole doping is $\delta=0.016$ ({squares}),
$\delta=0.049$ ({circles}), $\delta=0.093$ ({triangles}), and $\mu=0.58, 0.53$
and $0.48$ respectively. Solid line is the same quantity from ED-DMRG.
The ED-DMRG data are obtained for large finite clusters of 40 sites
and are plotted down to the frequency of the lowest energy pole.
Inset: inverse scattering time
Im$\Sigma_{pp}(\omega\rightarrow0)$ as a function of temperature.
Black symbols are for the same values of $\delta$ as in the main panel.
Open symbols are for higher doping $\delta = 0.44$ (circles)
and $0.62$ (squares) with $\mu=0.23$ and $0.13$ respectively.
Note that the inverse scattering time goes to $0$
in the Fermi liquid state.
}
\label{fig2}
\end{figure}

In the main panel of Fig.~\ref{fig1} we show the DOS of the metallic
state that results of lightly doping $\delta=3-n$ holes into
the ``parent'' Mott insulator state (shown in the right inset).
% which has a mix-valence character as $t_pd$ is sizable and we do not fix $n_d = 1$.
%The metallic nature of the doped state is evident from the finite
%spectral weight at $\omega=0$ in the DOS of both electronic components.
%The left inset of the figure shows a comparison of the QMC and ED-DMRG results
%for the DOS in a larger energy scale. The discrete peaks in the ED-DMRG data are
%due to the finite number of sites in the effective bath; nevertheless, the
%agreement in the distribution of spectral weights of the two methods is very
%satisfactory.
The main features of the metallic state are its mix-valence character and 
the presence of a quasiparticle peak at the Fermi energy that is 
flanked by lower and upper Hubbard bands \cite{sordi}. 
%Interestingly, it displays
%the celebrated peak-dip-hump lineshape often observed in photoemission studies
%of cuprate superconductors \cite{dessau}.
Within Fermi liquid theory, the key concept is that the one-particle excitations
near the Fermi energy, the ``quasiparticles'', are long lived entities. In consequence,
the imaginary part of their self-energy should go to zero as $\omega \to 0$.
More precisely, $\Sigma(i\omega_n) \approx i\omega_n a + b$ for small $\omega_n$,
with $a$ and $b$ real constants.
Quite surprisingly, however, the self-energy of the metallic
peak of Fig.~\ref{fig1} does not have this property.
In Fig.~\ref{fig2} we show the numerical results for
various paramagnetic metallic states obtained by small doping of the Mott
insulator.
%The metallic character is seen from the low frequency behavior of
%Im$G_{pp}(\omega_n)$, shown in the bottom inset.
The main panel shows 
%data for the
%imaginary part of the self-energy of the conduction electrons Im$\Sigma_{pp}$
%at the same parameter values.
the large finite intercept of Im$\Sigma_{pp}$ that reveals the NFL nature of the
states.
Therefore,
%In fact, the intercept provides an estimate of the inverse lifetime of the interacting
%one-particle states, so it indicates that 
the metallic peak at the Fermi energy of Fig.~\ref{fig1} is not due to quasiparticles
but rather to incoherent (\ie short lived) excitations.
In the lower
inset of Fig.~\ref{fig2} we show the extrapolated value of
Im$\Sigma_{pp}(\omega \to 0)$ that is an estimate of the
resistivity $\rho(T)$.
Contrary to what is expected for a metal, the resistivity increases with
decreasing $T$. At higher doping, however, it 
crosses over to conventional metallic (ie, Fermi liquid) behavior.
%the finite scattering time at low doping
%and its vanishing for higher dopings, as the system crosses over
%to the Fermi liquid.
This feature is in qualitative agreement with experiments 
in various NFL heavy fermion compounds \cite{stewart}.
%We should also note that with our QMC simulations we were able to  reach a
%very low ratio
The coherence temperature scale $T_{coh}$, below which the Fermi liquid behavior is obtained depends on $\delta$.
It is computed from the local spin susceptibility $\chi_{loc}$ and
is given by the x-axis intercept of $1/\chi_{loc}(T) \sim T+T_{coh}$. Physically, it
corresponds to the Kondo quenching of the local magnetic moments.
We found that $T_{coh}$ decreases with decreasing doping, and  becomes
vanishingly small at low $\delta$.
This feature is reminiscent
of the phenomenon of exhaustion, where few conduction electrons have to screen a large
number of spins. It has been extensively investigated in the Kondo lattice model \cite{burdin,costi}
and also in the PAM \cite{tjf1,tjf2,pruschke}. The exhaustion is usually realized in the limit 
where the number of local electrons is restricted
to unity, so to represent a spin at every lattice site, while the number of conduction
electrons is taken vanishingly small. Here, in contrast, as Fig.~\ref{fig1} shows, 
we are in a more mix-valence situation
where the number of $p$-carriers (holes) is not very small and never vanishes
(it is $\nu=0.14$ at $\delta=0$, and increases further with doping). 
Nevertheless, in the present case also the exhaustion physics is in fact responsible for the vanishing $T_{coh}$.
One should realize, however, that the relevant number of "conduction electrons" is not given
by the finite nominal $p$-hole occupation, but by $\delta$ which is the number of doped 
carriers to the (mix-valance) Mott insulator state.

Another important difference with respect to the standard exhaustion scenario is the stability
of the present NFL regime regarding magnetic ordering. The exhaustion regime is usually found to be 
unstable towards magnetic phases \cite{tjf1}. Here, in contrast we find that the NFL regime
is largely paramagnetic, except at very low dopings and temperatures, where the system becomes
an antiferromagnetic metal (AFM). The reason for the magnetic stability of the NFL regime can be argued 
to result from the magnetic frustration due to competing magnetic interactions. On one hand, neighboring
localized electrons in the Mott state interact via the superexchange mechanism, that follows from the
mapping of the narrow band to the Hubbard model. This interaction is antiferromagnetic
and is responsible for the AFM state that is found in the limit of very low hole doping. On the other hand, 
there is also a competing ferromagnetic interaction induced by hole doping. 
The doped holes need to delocalize to lower their kinetic energy, but they
are subject to a strong on-site magnetic binding to the local moments \cite{sordi}. Therefore,
in order to hop, they need the local moments of the neighbor sites to have the same
magnetic orientation as the one in its current site. 
These competing magnetic interactions lead to unquenched local dynamical fluctuations 
of the magnetic moments down to the vanishingly small $T_{coh}$, that provide the
source for the NFL scattering.

\begin{figure}%[!ht]
\centering
\includegraphics[width=9cm,clip]{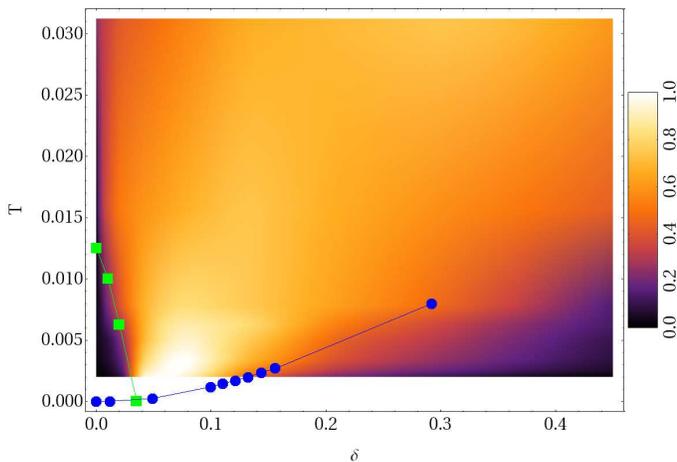}
\caption
{
Intensity plot of the scattering rate Im$\Sigma_{pp}(\omega \to 0)$
as a function of doping $\delta$ and $T$. 
For visualization, the scattering rates are normalized to max\{Im$\Sigma_{pp}(\omega_n)$\}
at each ($\delta$, $T$). The line with square dots at low
$\delta$ is $T_{Neel}$ and gives the AFM phase boundary (The $T$=0 data-point
is obtained from ED). The line with
circle dots denotes the numerical estimate for the
crossover scale $T_{coh}$. It is obtained from the extrapolation of
the low $T$ behavior of $1/\chi_{loc}(T)$ (see text).
}
\label{fig3}
\end{figure}

Fig.~\ref{fig3} condenses our results into a $\delta$-$T$ phase diagram.
The intensity plot shows the magnitude of the scattering rate Im$\Sigma_{pp}(\omega \to 0)$,
which is a measure of the NFL character of the system. We observe that conventional
metallic (ie, Fermi liquid) behavior occurs in two dark regions. The one at
low $\delta$ corresponds to the small AFM phase that we mentioned before, while the
one at higher $\delta$ is the Fermi liquid state that is realized beneath $T_{coh}$
due to conventional Kondo screening. In between these two regimes we find the
strong NFL behavior of the system.

\begin{figure}%[!ht]
\centering
\includegraphics[width=9cm,clip]{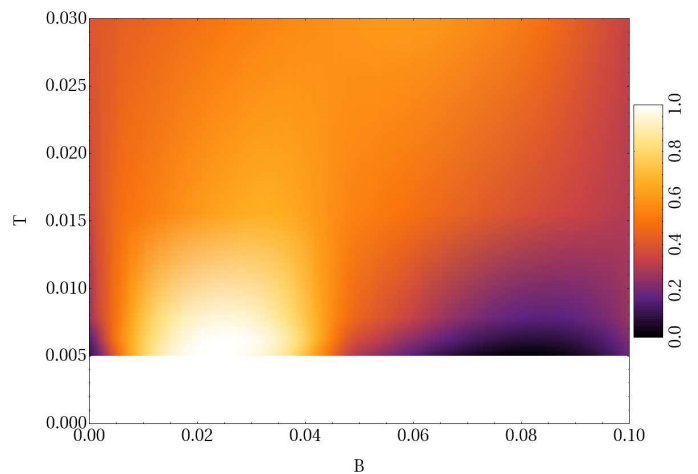}
\caption
{
Intensity plot of the scattering rate Im$\Sigma_{pp}(\omega \to 0)$
as a function of external magnetic field $B$ and $T$, at fixed doping $\delta = 0.01$. 
For visualization, the scattering rates are normalized to max\{Im$\Sigma_{pp}(\omega_n)$\}
at each ($B$, $T$).}
\label{fig4}
\end{figure}

The normal metal behavior in the AFM phase is easy to understand. 
As soon as the magnetic moments become N\'eel ordered 
the doped carriers have no problem to form coherent electronic waves. Thus, a
direct implication of this observation is that if one induces magnetic order,
the NFL state should become normal. To test this hypothesis 
we have mapped out the phase diagram as function of the external
magnetic field $B$ and $T$ at a small fixed doping. 
The results are shown in Fig.~\ref{fig4} which displays
the intensity plot of the computed scattering rate. At $B \to 0$ we find 
a small dark normal
metal region that corresponds to the AFM state discussed before. At large $B$ the
strong magnetic field induces ferromagnetic order of the local moments and, 
as expected, the metallic state is normal. Interestingly, in the intermediate $B$ region, 
a NFL regime emerges which can be understood as resulting from the competition between 
the magnetic interactions.

To conclude, we have studied the NFL regime that 
is realized upon doping a Mott insulator state in the Periodic Anderson model 
within the Dynamical Mean Field Theory.
The NFL
behavior is originated in the strong {\em local} magnetic scattering of the doped carriers
by unquenched fluctuating moments. This is in contrast to other approaches where the 
NFL state results from scattering by spatial magnetic fluctuations.  
We find that both, doping and magnetic field, allow to tune into the NFL regime, in
qualitative agreement with some of the non-Fermi liquid phenomenology observed in 
heavy fermion systems.

We acknowledge M.Gabay, E.Miranda and D.J.Garc\'ia for useful discussions.
AA acknowledges support from the European ESRT Marie-Curie program,
GS and MJR from the ECOS-Sud program.


\begin{thebibliography}{99}
\bibitem{note}
$p$ and $d$ orbitals may also denoted $c$ and $f$ respectively.
\end{thebibliography}
\end{document}